\documentclass[reprint,aps,superscriptaddress,floatfix,showpacs]{revtex4-2}
\usepackage[utf8]{inputenc}
\usepackage{amsmath}
\usepackage{amssymb}
\usepackage{graphicx}

\makeatletter
\usepackage{graphics}
\usepackage{epsfig}
\usepackage{color}
\newcommand{\be}{\begin{equation}} \newcommand{\ee}{\end{equation}}
\newcommand{\bea}{\begin{eqnarray}} \newcommand{\eea}{\end{eqnarray}}

\makeatother

\usepackage{xcolor}

\begin{document}

\title{Nearly Hamiltonian dynamics of laser systems}

\author{Antonio Politi}
\affiliation{Institute of Pure and Applied Mathematics, Department of Physics, Aberdeen AB24 3UE, United Kingdom}
\author{Serhiy Yanchuk} 
\affiliation{Potsdam Institute for Climate Impact Research, Telegrafenberg A 31, 14473 Potsdam, Germany}
\affiliation{Institute of Mathematics, Humboldt University Berlin, 12489 Berlin, Germany}
\author{Giovanni Giacomelli} 
\affiliation{Consiglio Nazionale delle Ricerche, Istituto dei Sistemi Complessi, via Madonna del Piano 10, I-50019 Sesto Fiorentino (FI), Italy}

\date{\today}

\begin{abstract}
The Arecchi-Bonifacio (or Maxwell-Bloch) model is the benchmark for the description of active optical media. 
However, in the presence of a fast relaxation of the atomic polarization, its implementation is a challenging
task even in the simple ring-laser configuration, due to the presence of multiple time scales.
In this Article we show that the dynamics is nearly Hamiltonian over time scales much longer 
than those of the cavity losses.
More precisely, we prove that it can be represented as a pseudo spatio-temporal pattern generated by a
nonlinear wave equation equipped with a Toda potential. 
The existence of two constants of motion (identified as pseudo energies), thereby, 
elucidates the reason why it is so hard to simplify the original model:
the adiabatic elimination of the polarization must be accurate enough to describe the dynamics correctly over 
unexpectedly long time scales.
Finally, since the nonlinear wave equation with Toda potential can be simulated on much longer times than the previous
models, this opens up the route to the numerical (and theoretical) investigation of realistic setups.
\end{abstract}

\maketitle

Optical active media are of fundamental importance in several fields, ranging from signal enhancement in detection 
setups~\cite{Frede2007} to regeneration of digital transmissions in fibers \cite{Li2017}, and chirped 
pulse amplification~\cite{Strickland1985}. The most appropriate model to analyse such phenomena are the Arecchi-Bonifacio~\cite{Arecchi1965} (AB) equations
(often called Maxwell-Bloch~\cite{Mcneil2015}),
derived from first principles, under the slowly varying envelope approximation to remove the
optical frequencies. The most important application is by far in lasers, where the 
coherent amplification process produces a strong emission of radiation with striking properties as observed e.g. in 
fiber systems~\cite{Turitsyna2013,Rogers2005,Abarbanel1999,Ray2008}. 
The model has been extensively studied in unidirectional ring lasers for more than 50 years, allowing to determine the first and second
laser threshold  in the single-mode description, the Risken-Nummedal-Graham-Haken 
(RNGH)~\cite{Risken1968,Risken1968a,Graham1968} instability in the full multi-mode model, and to characterize the emergent chaotic dynamics (see e.g., 
\cite{Lugiato2018a,Gorni2019,Roldan2001,Roldan2003,Lugiato1986,Menegozzi1973,Haken1975,BCL,Lugiato2010,
Milonni1987,Elgin1987,Ikeda1989}).

The computation time can, however, be exceedingly long in the presence of a comparably fast polarization dynamics:
a rather common feature encountered in light-amplifying devices~\cite{Bellemare2003a,Choudhury2018} and 
optical networks ~\cite{Giacomelli2019,Giacomelli2020,Soriano2013,Huang2022}.
It is, therefore of paramount importance to overcome the difficulty posed by the presence of multiple time scales. 
Unfortunately, in spite of its extremely fast relaxation time, the atomic-polarization variable cannot 
be plainly adiabatically eliminated, since the resulting model exhibits unphysical divergencies. 
In mode-locking setups, the problem is not perceived, since they are typically based on the Haus Master equation~\cite{Haus1975}.
This is a phenomenological equation which does not include the atomic variables,
consequently bypassing the unpleasant instabilities. 
Moreover, the Haus equation does not only lack a first-principle justification; 
it is also inaccurate in certain important cases: as noted in~\cite{Perego2020}, the RNGH instability is not reproduced 
in the case of a plain ring laser, even when an equation for the population is added a posteriori. 
In fact, a more rigorous strategy requires a refined adiabatic
elimination, as for instance proposed in~\cite{DeValcarcel2003,Perego2020}.  
In this Article, we follow a similar strategy with a couple of differences:
we start from the simpler and yet accurate ``spaceless" delayed representation~\cite{Giacomelli2021}
with no periodic intra-cavity modulations. 
Time-delayed models had been already proposed for both the unidirectional propagation in a ring laser 
\cite{Vladimirov2005} and mode-locked lasers \cite{Schelte2020}. However, they do not rigorously account 
for the atomic variables, as done in Ref.~\cite{Giacomelli2021}, which instead includes
the polarization dynamics.
By implementing a clean perturbative approach, the spaceless delayed model is progressively simplified,
finally leading to a
nonlinear wave equation 
with Toda potential (NWT). 
This result generalizes the observation made in the '80s that a single-mode laser 
characterized by a large polarization decay-rate can be viewed as a slowly relaxing Toda oscillator~\cite{Oppo1985,Oppo1986}.
The extension is nontrivial since the Hamiltonian here involves an infinity of modes
and, furthermore, the  ``space" variable is actually a mixture of different time-scales.
This latter property reveals also a substantial difference with the early work by Haken to describe
multi-mode lasers \cite{Haken1977}, where a ``free" energy was derived in the standard space-time representation.
Last but not least, it is remarkable that the NWT model is to a large extent valid independently of the population
decay rate and of the cavity losses which enter only to define the effective length and, indirectly, the corresponding
NWT energy.

We start by briefly recalling the AB equations, since they are the fundamental reference: this is the model which takes into account
the atomic-variable dynamics as from the theory of two-level atoms. 
For the sake of simplicity, we restrict ourselves to the resonant case and thus assume that the atomic
polarization $\mathcal{P}$ and the electric field $\mathcal{F}$ are real, as well as the population inversion
$\mathcal{D}$.
In a comoving frame, the model can be written as,~\cite{Giacomelli2021}
\begin{eqnarray}
\frac{\partial\mathcal{F}}{\partial y}  &=& \frac{a}{2}\mathcal{P},  \nonumber \\
\frac{\partial\mathcal{P}}{\partial \hat t} &=& \mathcal{D}\mathcal{F}-\mathcal{P}, \label{MB}\\
\frac{\partial \mathcal{D}}{\partial \hat t} &=& \gamma \left [1-\mathcal{D} - \mathcal{F}\mathcal{P})\right ]~, \nonumber
\end{eqnarray}
where $y$ is the scaled spatial variable ($y\in[-1,1]$).
The parameter $\gamma$ is the ratio $\gamma_\parallel/\gamma_\perp$, where $\gamma_\parallel$ and $\gamma_\perp$ are the 
population and polarization decay rates, respectively,
The time $\hat{t}$ is expressed in units of $\gamma_\perp^{-1}$, while
$a$ is the pump parameter, controlling the amount of energy injected into the laser. 
The boundary condition is $\mathcal{F}(y=-1,t) = R \mathcal{F}(y=1,t-\mathcal{T})$,
$R$ being the mirror reflectivity and $\mathcal{T}$ the round-trip time across the cavity.

In Ref.~\cite{Giacomelli2021}, we have shown that this set of equations is generally well approximated by a spaceless delayed model,
\begin{eqnarray}
F(\hat t) &=& R F(\hat t-\mathcal{T}) + a P(\hat t) \nonumber, \\
\frac{\mathrm{d} P(\hat t)}{\mathrm{d} \hat t}  &=& -P(\hat t) + D(\hat t)F(\hat t),  \label{eq:ABdelay} \\
\frac{\mathrm{d} D(\hat t)}{\mathrm{d} \hat t}  &=& \gamma(1-D(\hat t)-F(\hat t)P(\hat t)) \; . \nonumber
\end{eqnarray}
This is a slightly simplified version of the equations derived in \cite{Giacomelli2021}; we have, however,
verified that they reproduce the regime generated by the original AB model 
(see below the discussion of numerical simulations).
Here, $P(\hat t) = \langle \mathcal{P}(y,\hat t) \rangle$,
$D = \langle  \mathcal{D}(y,\hat t) \rangle$ (the angular brackets denote a spatial average),
and finally $F(\hat t) = \mathcal{F}(1,\hat t)$.

In a large fraction of laser devices, $\gamma\ll 1$; these are the so-called class-B lasers. 
From now on, we consider this class of systems.
It is well known that although the polarization relaxes quickly, its adiabatic
elimination does not lead to a meaningful model~\cite{DeValcarcel2003}.
It is convenient to rescale the time, defining $t = \Gamma \hat t$, where $\Gamma = \sqrt{\gamma}$
(accordingly $T = \Gamma \mathcal{T}$), and to introduce the rescaled pump-to-losses parameter
\begin{equation}
	I = \frac{a}{1-R} -1  \; .
\end{equation}
Next, by following an observation made in Ref.~\cite{DeValcarcel2003} about the amplitude of the oscillations of
$P$ and $D$ around their equilibrium value,
we introduce a new set of variables, namely
$f = F/\sqrt{I}$, $u = \big( P(I+1)-F \big)/(\Gamma\sqrt{I})$, and $g = \big( D(I+1)-1 \big)/\Gamma$. 
As a result, model (\ref{eq:ABdelay}) can be rewritten as
\begin{eqnarray}
f(t) &=& f(t-T)+ \Gamma \frac{1-R}{R}u(t), \nonumber \\
 \frac{\mathrm{d} u(t)}{\mathrm{d}t} &=& \frac{R}{\Gamma} \left[-u(t) + g(t)f(t) -  \frac{\mathrm{d} }{\mathrm{d}t} f(t-T)\right], \label{eq:scaled}\\
 \frac{\mathrm{d} g(t)}{\mathrm{d}t} &=& -\Gamma g(t) + I(1- f^2(t) -\Gamma f(t)u(t)). \nonumber
\end{eqnarray}
We now adiabatically eliminate $u$ by setting $\dot u = 0$,
\begin{equation}
u(t) = g(t)f(t) -\frac{\mathrm{d} }{\mathrm{d}t}  f(t-T) \; .
	\label{eq:elim}
\end{equation}
Although this relationship is more accurate than the one obtained by setting the time derivative of $P$ equal to 0,
(we would have missed the last term in Eq.~(\ref{eq:elim})), it is still not sufficiently accurate 
to reproduce all the relevant details.
However, Eq.~(\ref{eq:elim}) is an important step forwards as it allows understanding the origin
of the singularity of the limit $\Gamma \to 0$, thereby
identifying the backbone of the dynamical evolution.
By plugging Eq.~(\ref{eq:elim}) into Eq.~(\ref{eq:scaled}), we obtain
\begin{eqnarray}
	f(t) &=& f(t-T) +  \varepsilon\left( g(t)f(t) - \frac{\mathrm{d} }{\mathrm{d}t} f(t-T)\right), \label{eq:fdel}\\
 \frac{\mathrm{d} g(t)}{\mathrm{d}t} &=& I\left(1- f^2(t)\right) +\mathcal{O}(\varepsilon ), \label{eq:geq}
\end{eqnarray}
where we have introduced the smallness parameter
\begin{equation}
\varepsilon = \frac{1-R}{R}\Gamma~,
\end{equation}
and leave the correction terms in Eq.~(\ref{eq:geq}) unspecified. Later we show that they are not relevant for the
identification of the leading order of the dynamics.

We now transform the dependence on the single variable $t$ into a spatio-temporal representation. This can be done by formally using the multiscale approach, which introduces slow and fast timescales $\varepsilon t$ and $t$, as described in \cite{Wolfrum2006,Yanchuk2014,Yanchuk2017}. 
In such a case, the fast timescale plays the role of space and the slow timescale plays the role of time. 
However, we proceed here with a more phenomenological approach as in \cite{Giacomelli1996} (leading to the same result) by
introducing $(\zeta,\xi)$, i.e., $f(t)=\bar f(\zeta,\xi)$ and $g(t)=\bar g(\zeta,\xi)$,
where $\zeta = t \! \mod T$ is a pseudo-spatial variable, while $\xi =t/T$ is a 
pseudo-temporal variable. 
This transformation can be intuitively interpreted as ``wrapping" the time axis around
a cylinder of circumference $T$ in such a way
that the longitudinal coordinate (the new time) increases by one unit after a full rotation -- see more details in \cite{Giacomelli1994,Giacomelli1996,Yanchuk2014,Yanchuk2017}. 
It is easily seen that periodic boundary conditions hold in the space-time representation:
$\bar f(\zeta=0,\xi) = \bar  f(\zeta=T,\xi)$ (analogously for $\bar g$).
Although this transformation was originally introduced and mainly used in the context of long delays, it is justified for arbitrary delay, like here.

This transformation is very useful since,
from Eq.~(\ref{eq:fdel}) we see that $\bar f(\zeta,\xi) - \bar f(\zeta,\xi-1) = \mathcal O(\varepsilon)$.
Hence, it is legitimate to expand $\bar f(\zeta,\xi-1)$ around $(\zeta,\xi)$. If, moreover,
we rescale the time axis, introducing, 
$\theta = \varepsilon \xi$ we can write,
\begin{equation}
f(t-T) = \bar f(\zeta,\xi-1) = \bar f(\zeta, \theta-\varepsilon)\approx \bar f (\zeta, \theta) -\varepsilon \bar f_\theta (\zeta, \theta),
\end{equation}
where the subscript denotes a derivative with respect to the corresponding variable. The change of
variables from $t$ to $(\zeta,\theta)$ implies also that
\begin{equation}
\frac{\mathrm{d}f}{\mathrm{d} t} = \bar f_\zeta + \frac{\varepsilon}{T} \bar f_\theta .
\end{equation}
By substituting this into Eq.~(\ref{eq:fdel}) we obtain
\begin{equation}
	\bar f_\theta + \bar f_\zeta =  \bar g \bar f,  \label{eq:1a}
\end{equation}
where we have neglected quadratic and higher order terms in $\varepsilon$. 
As terms of order $\mathcal{O}(\varepsilon)$ are absent in the first equation, we are entitled to
neglect them in Eq.~(\ref{eq:geq}) too. As a result we obtain
\begin{equation}
\bar g_\zeta = I\left(1- \bar f\, ^2  \right).  \label{eq:1b}
\end{equation}
Eqs.~(\ref{eq:1a},\ref{eq:1b}) constitute the minimal description of the laser dynamics in the 
small $\Gamma$ limit. The model is completed by recalling that the two fields satisfy
periodic boundary conditions on the domain of size $\zeta_L = T$.
A further, non-trivial simplification can be achieved by introducing
$q = \bar g/\sqrt{I}$, $s = \ln \bar f$, and rescaling the variables as 
$\sigma = \zeta\sqrt{I}$, $\tau = \theta\sqrt{I}$.
As a result, the explicit dependence on  $I$ is removed,
\begin{equation}
s_\tau = -s_\sigma + q \quad ,\quad
q_\sigma = 1- e^{2s}  \; ,
\label{eq:KGT0}
\end{equation}
and it appears only in the definition of the ``spatial" length $L =T\sqrt{I}$.
This equation can be given a simple physical interpretation by 
introducing the new time-like variable $y= (\tau+\sigma)/2$ and 
the new space-like variable $x = (3\tau-\sigma)/2$. In fact,
the model can be written as a single second-order PDE
\begin{equation}
	\label{eq:KGT}
s_{yy} - s_{xx} = 2(1- e^{2s}) \equiv  -V_s(s)~,
\end{equation}
(the subscript denotes a derivative with respect to the variable $s$).
This is a nonlinear wave equation 
equipped with a Toda potential $V(s) = \textrm{e}^{2s}-2s-1$~\cite{Toda1975}.
Notice that the validity of the Hamiltonian model depends only on the smallness
of $\varepsilon$, irrespective of the presence of finite cavity losses ($R < 1$).

Equation~(\ref{eq:KGT}) generates a Hamiltonian 
dynamics.  Being the NWT model translationally invariant in space, we expect 
energy and impulse to be both conserved.
In the more physical $(\sigma,\tau)$ description,
the simplest independent conservation laws can be formulated 
in terms of the two densities
\begin{equation}
  h_K =  \frac{s_\sigma^2}{2} \quad , \quad
  h_P = V(s) \; .
\end{equation}
With the help Eq.~(\ref{eq:KGT0}) and recalling the definition of $V(s)$,
one can prove that the integral of $h_K$ is constant,
\begin{equation}
\partial_\tau\!\!  \int_0^L d\sigma h_K =
\int_0^L d\sigma~ s_\sigma \!\left( - s_{\sigma\sigma} + V_s \right) = 0 \; .
\label{cons1}
\end{equation}
where we have exploited the periodic boundary conditions.
Analogously,
\begin{eqnarray}
\partial_\tau\!\!  \int_0^L d\sigma h_P &=&
\int_0^L\!\! d\sigma~s_\tau V_s(s) = \int_0^L \!\!d\sigma~\left(-s_\sigma +q\right)V_s(s) = \nonumber \\
&&
-\int_0^L \!\!d\sigma~s_\sigma V_s(s) - 2\int_0^L d\sigma~q q_\sigma = 0 \, . 
\label{cons2}
\end{eqnarray}
By recalling that $\sigma$ corresponds to the original ``fast" time scale, 
$h_K$ can be read as a ``kinetic'' energy (KE) density.
On the other hand, $h_P$ can be interpreted as a density of ``potential" energy (PE),
which accounts for the field fluctuations weighted according to the Toda potential.
The existence of these two conservation laws is the primary reason why 
the adiabatic elimination of the polarization is such a delicate problem:
the leading terms controlling the dynamics of the KE and PE must be captured
correctly, and they require going one order beyond in the perturbation
expansion.

In the limit $\Gamma \ll 1$, the NWT model is the backbone of an amplitude equation for the laser dynamics. 
In a sense, this work complements the analysis of Ref.~\cite{Casini1997} performed for arbitrary $\Gamma$, but restricted
to the region close to the RNGH threshold, where dynamical spatial structures emerge, 
and explains the singularity of the normal-form 
therein derived as a manifestaton of vanishing losses. 
Being the NWT model Hamiltonian, it is not structurally stable:
arbitrarily small $\Gamma$ generically brings in dissipation and amplification, 
via the herein neglected higher-order terms.
Such terms affect the temporal evolution of the energy densities KE and PE, which are no longer constant.
Nevertheless, over time-scales smaller than $\tau_l \approx 1/(\sqrt{I}\varepsilon)$ perturbations should be negligible
and the energies thereby conserved.

\begin{figure}
\includegraphics[width=0.5\textwidth,clip=true]{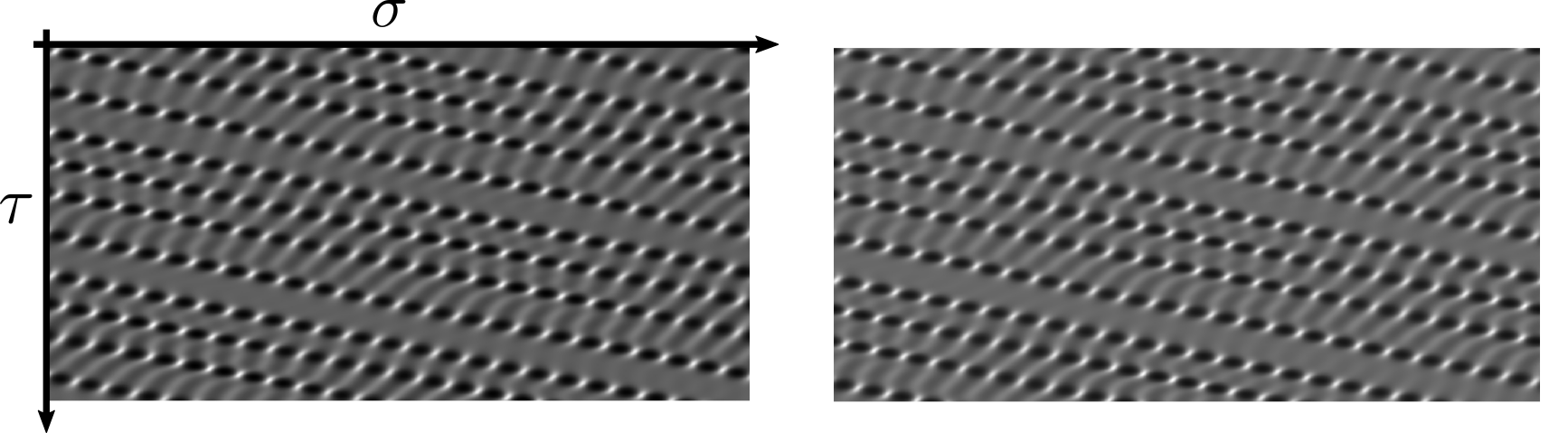}
\caption{Spatio-temporal representation of the logarithm of the field amplitude in grayscale, from black (0.2) to white (2.3), as obtained from the integration of AB model (\ref{MB}) (left) and Eqs.~(\ref{eq:ABdelay}) (right).  
Parameters are: $a=7$, $R=0.95$, round-trip time $\mathcal{T}$ equivalent to 
$L= 10 \sqrt{139} \approx 117.9$, and $\gamma = 10^{-4}$ ($\Gamma=10^{-2}$). 
$2\times10^4$ cells (delays) are shown, corresponding to $\tau=124.1$ time unit.
The initial condition for the delayed model has been fixed by setting
$P$ and $D$ equal to their value at the end of the cavity in the AB equations.
\label{fig:pattABvsDEL} 
}
\end{figure}

We now start validating he spaceless model~(\ref{eq:ABdelay}) by comparing it with the original AB
equations~{\bf \ref{MB}}.
In Fig.~\ref{fig:pattABvsDEL} the AB pattern generated after discarding a long transient is plotted (left) alongside
with the outcome of the corresponding spaceless model (right). The pump value $a=7$ is much above the RNGH threshold 
(linear stability shows that the instability arises above $a\approx 0.45$ \cite{Lugiato1986})
and long enough delay ($\tau=124.1$) to yield a multi-mode dynamics as also testified by
the irregular spatial structure.
It is transparent that the delayed model provides a fairly accurate
representation of the laser dynamics and can, thereby, be considered as a reliable starting point.

\begin{figure}
\includegraphics[width=0.5\textwidth,clip=true]{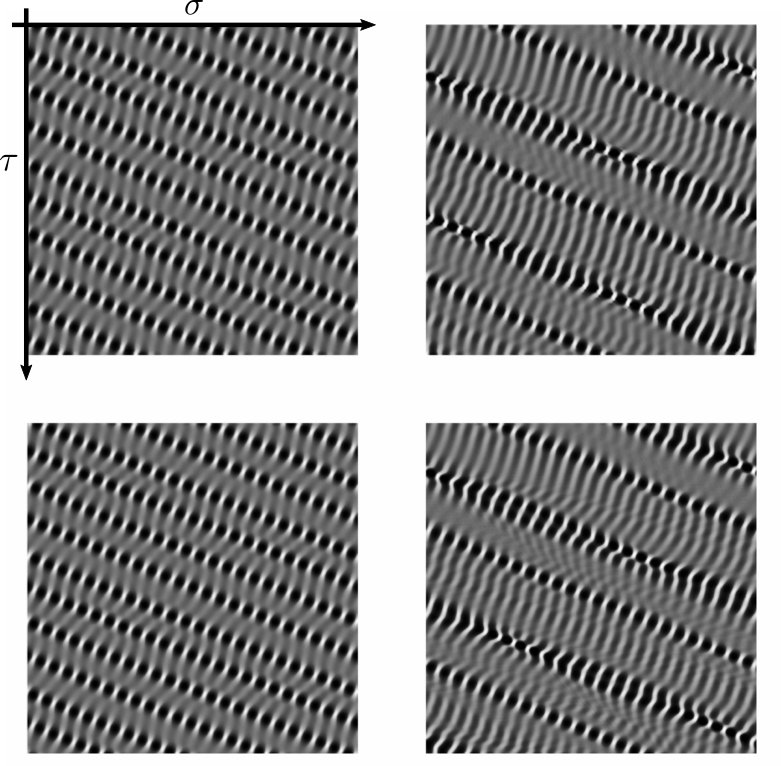}
\caption{
\label{fig:patt} 
Spatio-temporal representation of the logarithm of the field amplitude, in grayscale from black (0.2) to white (2.3) for the left column, and from black (0.1) to white (3.0) for the right column. The two upper panels are obtained by integrating model (\ref{eq:ABdelay}); the lower ones by integrating (\ref{eq:KGT}).
	Left and right columns correspond to two different initial conditions. Parameters are as in Fig.\ref{fig:pattABvsDEL}, except for  $\gamma = 10^{-4}/64$ ($\Gamma = 0.00125$) and a total of $\tau=117.9$ effective time units.}
\end{figure}

Next, we compare the spaceless model with the Hamiltonian one~(\ref{eq:KGT0}).
The delayed equations (\ref{eq:ABdelay}) have been simulated for the parameters reported in Fig.~\ref{fig:pattABvsDEL}, with $\gamma = 10^{-4}/64$ ($\Gamma = 0.00125$)~\footnote{This is the smallest $\Gamma$ value we 
could numerically afford. Though small, it is still much larger than what expected in e.g. Erbium lasers,
where $\Gamma \approx 10^{-5}$~\cite{Pessina1997}}.
We have selected a couple of instantaneous configurations in the stationary regime and used them as initial
conditions to generate spatio-temporal patterns.
The outcome is presented in the two upper panels of
Fig.~\ref{fig:patt}.
We have then simulated the NWT model, rewriting it as, 
\begin{equation}
\label{approx3}
	q_{\sigma\tau} = -q_{\sigma\sigma} - 2q(1-q_\sigma) \; .
\end{equation}
This equation can be seen as a first order PDE for the field $q_\sigma(\sigma,\tau)$, where
$q(\sigma,\tau)$ is determined via a spatial integration, under the condition that the spatial average of $q$ is zero, as
required by the boundary conditions.

The NWT equation has been integrated by using a fourth-order finite-difference algorithm, with  a time step $\delta = 0.005$
and $N=8000$ points to discretize the space.
The resulting patterns, obtained by starting from the same initial conditions,  
are presented in the two lower panels of Fig.~\ref{fig:patt}.
The excellent agreement confirms the Hamiltonian-like nature of the underlying dynamics.

Since direct numerical simulations display a high-frequency instability, we have performed a spatial
smoothing of amplitude $10^{-4}$ every 100 time steps, which does not affect the conservation laws over the explored time scales.
Whether this instability is the consequence of an ill posed Cauchy problem or not, it is immaterial, since higher-order
perturbative terms are nevertheless present, and the overall stability should be judged after including such
corrections. This will be the task of future work.

\begin{figure}
	\includegraphics[width=0.5\textwidth,clip=true]{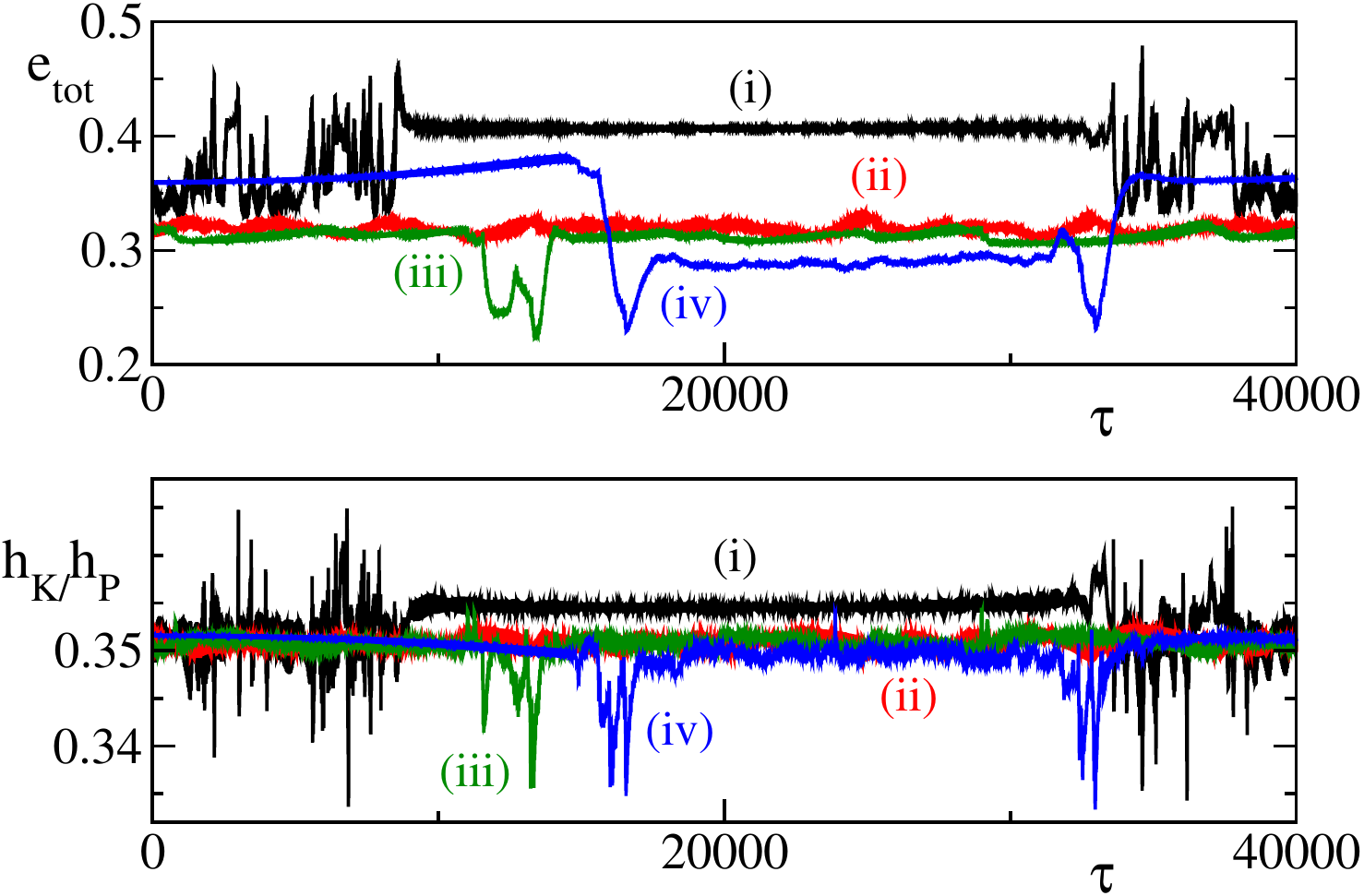}
\caption{Total energy $e_\mathrm{tot} = h_K  + h_P$ (upper panel) and the ratio between the kinetic and potential energies $h_K/h_P$ (lower panel) from direct integration of system \eqref{eq:ABdelay} (see text). The colors refer to $\Gamma$ values of (i) 0.01; (ii) 0.005; (iii) 0.0025; and (iv) 0.00125.
	\label{fig:energy} }
\end{figure}

Next, we have tested the behavior of the delayed model over yet longer time scales.
The results obtained for different $\Gamma$ values are reported Fig.~\ref{fig:energy}.
In the upper panel, the total energy density $e_\mathrm{tot} = h_K  + h_P$ 
is plotted vs. time (the other parameters are as in Fig.~\ref{fig:patt}).
There, we see that $e_\mathrm{tot}$ is approximately constant and independent of $\Gamma$.
The temporal fuctuations are a consequence of the neglected terms: 
the few jumps are quite likely induced by pattern selection mechanisms.

Interestingly, the smallness of the self-selected energy density ($\approx 0.3/0.4$) is suggestive of weak chaos,
since in the linear limit with vanishing energy density the model is obviously integrable. 
This is consistent with
the regularity of the patterns displayed in Figs.~\ref{fig:pattABvsDEL},\ref{fig:patt}.

In the lower panel of Fig.~\ref{fig:energy}, we plot the ratio between KE and PE. It is nearly constant and close to 0.35. 
We have verified that the same holds true in a relatively broad range 
of $a$-pump values from 1 to 8. In equilibrium statistical mechanics, the ratio is 1. A refinement of our theory,
including higher-order terms is necessary to explain the deviation.

In conclusion, we have shown that the ring-laser dynamics is well reproduced by the NWT model over
the fast ``spatial" scale $t_f \approx (\gamma_\parallel \gamma_\perp)^{-1/2}$,
and the longer Hamiltonian scale $\tau \approx 1$, which corresponds to
$t_H \approx  \mathcal{T}_p R/[\Gamma \sqrt{a(1-R)}]$, where $\mathcal{T}_p$ is the round trip time (all in physical units).
Moreover, from the size of the neglected terms, we can predict
the yet longer time scale $t_l \approx t_H \tau_l \approx \mathcal{T}_p/[\Gamma^2 a(1-R)]$, which is also
the scale over which the two energies are expected to fluctuate. A fully quantitative analysis requires going one order
beyond in the perturbative analysis: we leave this task to future work.
The separation of time scales is better appreciated by considering  
a common Erbium laser configuration \cite{Pessina1997}.
Referring to effective two-level-system parameters 
(see, e.g., \cite{BCL}) $\gamma_\perp = 2\pi\times10^{12} \textrm{s}^{-1}$, 
$\gamma_\parallel = 5\times 10^3 \textrm{s}^{-1}$ and $\mathcal{T}_p=100$~ns, together with $a=7$ 
and $R=0.95$ here considered, we obtain $t_f \approx  5.6$~ns, $t_H \approx 5.7$~ms  and $t_l = 3.6$~s, with a span of almost 9 orders of magnitude. 
Notice that $t_H$ is much longer than the inverse of the field decay rate ($\approx 10 ~\mu$s), 
showing that the Hamiltonian character occurs over time scales when 
the field dynamics has been fully affected by its dissipative losses.

Remarkably, the Hamiltonian conservation laws (\ref{cons1},\ref{cons2}) can be interpreted as the signature of a marginal stability of the system at this level of expansion. As a consequence, we have provided a rigorous ground to the common habit to move at least to the second order in the adiabatic elimination.     

Finally, a further important advantage of the NWT equation: the presence of the much compressed time scale (the
variable $\tau$) allows
integrating the dynamical equations by using a time step much longer (by 3-4 orders of magnitude) than in the best
current model (the PDE derived in Ref.~\cite{Perego2020}).

It is instructive to compare our results with those in Ref.~\cite{Seidel2022}, 
where light propagation is investigated in a Kerr medium.
In that paper too, the authors start from
a delay-algebraic system, to then derive a Hamiltonian amplitude-equation, which, in their case, has the form of
a nonlinear Schr\"odinger equation.
The similarity of the conclusions suggests that delay-algebraic models may possess general, still unearthed, properties.
In fact, in spite of the analogies, the two models are substantially different:
unlike in Ref.~\cite{Seidel2022}, our discussion of the propagation 
takes into account the role of polarization and does not require the long delay limit.

Our results pave the way to a series of routes that should be explored. 
Distributed losses arising from light propagation are often negligible; to
what extent, their inclusion can change the overall scenario? 
Is the nearly Hamiltonian representation still valid in the presence of detuning as it happens 
in the single mode case~\cite{Oppo89}?
Finally, what happens in semiconductor lasers, where the pump parameter $a$ needs to be multiplied by a complex 
phase factor~\cite{Henry1982}?
Minor qualitative changes are expected (see e.g. \cite{Perego2020}), but the conclusion needs to be validated.

\begin{acknowledgments}
One of us (AP) wishes to thank Vieri Benci for enlightening discussion on nonlinear Klein-Gordon dynamics. The work of S.Y. was supported by the German Research Foundation DFG, Project
No. 411803875.
\end{acknowledgments}

%

\end{document}